\documentclass[sigconf]{acmart}
\usepackage{multirow}
\usepackage{ifpdf}
\AtBeginDocument{%
	}

\copyrightyear{2025} 
\acmYear{2025} 
\setcopyright{acmlicensed}\acmConference[RecSys '25]{Proceedings of the
	Nineteenth ACM Conference on Recommender Systems}{September 22--26,
	2025}{Prague, Czech Republic}
\acmBooktitle{Proceedings of the Nineteenth ACM Conference on
	Recommender Systems (RecSys '25), September 22--26, 2025, Prague, Czech
	Republic}
\acmDOI{10.1145/3705328.3748091}
\acmISBN{979-8-4007-1364-4/2025/09}

\begin{document}
	
	\title{IP2: Entity-Guided Interest Probing for Personalized News Recommendation}
	
	\author{Youlin Wu}
	\affiliation{
		\institution{Dalian University of Technology}
		\city{Dalian}
		\country{China}
	}
	\email{wuyoulin@mail.dlut.edu.cn}
	
	\author{Yuanyuan Sun}
	\affiliation{
		\institution{Dalian University of Technology}
		\city{Dalian}
		\country{China}
	}
	\email{syuan@dlut.edu.cn}
	
	\author{Xiaokun Zhang}
	\authornote{Corresponding author.}
	\affiliation{
		\institution{City University of Hong Kong}
		\city{Hong Kong}
		\country{Hong Kong}
	}
	\email{dawnkun1993@gmail.com}
	
	\author{Haoxi Zhan}
	\affiliation{
		\institution{Dalian University of Technology}
		\city{Dalian}
		\country{China}
	}
	\email{zhanhaoxi@mail.dlut.edu.cn}
	
	\author{Bo Xu}
	\affiliation{
		\institution{Dalian University of Technology}
		\city{Dalian}
		\country{China}
	}
	\email{xubo@dlut.edu.cn}
	
	\author{Liang Yang}
	\affiliation{
		\institution{Dalian University of Technology}
		\city{Dalian}
		\country{China}
	}
	\email{liang@dlut.edu.cn}
	
	\author{Hongfei Lin}
	\affiliation{
		\institution{Dalian University of Technology}
		\city{Dalian}
		\country{China}
	}
	\email{hflin@dlut.edu.cn}
	
	
	\begin{abstract}
		News recommender systems aim to provide personalized news reading experiences for users based on their reading history. 
		Behavioral science studies suggest that screen-based news reading contains three successive steps: scanning, title reading, and then clicking.
		Adhering to these steps, we find that intra-news entity interest dominates the scanning stage, while the inter-news entity interest guides title reading and influences click decisions. Unfortunately, current methods overlook the unique utility of entities in news recommendation. To this end, we propose a novel method called IP2 to probe entity-guided reading interest at both intra- and inter-news levels. 
		At the intra-news level, a Transformer-based entity encoder is devised to aggregate mentioned entities in the news title into one signature entity. 
		Then, a signature entity-title contrastive pre-training is adopted to initialize entities with proper meanings using the news story context, which in the meantime facilitates us to probe for intra-news entity interest. 
		As for the inter-news level, a dual tower user encoder is presented to capture inter-news reading interest from both the title meaning and entity sides. 
		In addition to highlighting the contribution of inter-news entity guidance, a cross-tower attention link is adopted to calibrate title reading interest using inter-news entity interest, thus further aligning with real-world behavior.
		Extensive experiments on two real-world datasets demonstrate that our IP2 achieves state-of-the-art performance in news recommendation.
	\end{abstract}

	
	
	\maketitle
	
	\section{Introduction}
	The rapid expansion of the internet provides users with convenient access to a vast amount of news stories from various online media platforms. However, this abundance also presents a significant challenge of information overload, making it difficult for users to find news items that align with their interest \cite{wu-etal-2020-mind}.
	To address this, news recommender (NR) systems have been widely adopted to personalize news delivery, thereby enhancing user engagement and satisfaction \citep{li2019survey, jeng2024bridging}.
	Unlike commonly encountered commodity recommendation \cite{zhang2024finerec}, news articles are characterized by their rich semantic contents and time-sensitive nature, conveying significant events related to various entities such as people and places \citep{fenton2009news, feng2020news}. This inherent complexity of news stories necessitates careful consideration of how to effectively leverage these rich information sources within the recommendation process.
	\label{intro:entity}
	\begin{figure}[t]
		\centering
		\includegraphics[width=0.95\columnwidth]{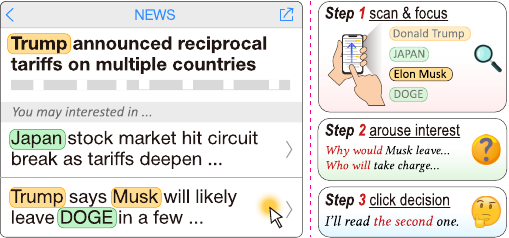} 
		\caption{An example of a complete news selection chain in news reading. A reader first spots ``Elon Musk'' during \textit{scanning}. Then, after intensive \textit{title reading}, they raise questions that ultimately lead to a \textit{click} on the second news entry.}
		\label{fig.intro}
	\end{figure}
	
	Advancements in deep learning have propelled methods based on neural collaborative filtering (NCF) \cite{he2017neural, rendle2020neural, zhang2025survey} to the forefront of news recommendation. These methods utilize semantic features derived from news content (\textit{e.g.,} titles) to model user preferences. To achieve this, they employ news and user encoders to learn corresponding embeddings. Subsequently, the probability of a user selecting one news item is determined based on embedding similarity \cite{wu2019NRMS, wu2021PLMNR, Qi2022FUMFA}. 
	Beyond regular NCF, leveraging external knowledge has become a promising approach for a deeper understanding of reading preferences. Two main lines of research have gained significant attention. 
	One line of work focuses on enhancing news representations by integrating entity embeddings learned from knowledge graphs \cite{wang2018dkn, qi2021KIM, liu2020kred}. 
	The other line of work focuses on incorporating entities into the user encoder to capture the intricate connections between them, thereby achieving a more fine-grained model of user reading preferences \cite{qiu2022graph, yang2023GLORY}. 
	
	Although these methods have greatly advanced news recommendation, they only treat entities as a complement to the semantic features, without noticing the unique role played by the entity itself.
	Behavioral science studies \cite{liu2005reading} reveal that people tend to take more time on \textit{scanning, keywords spotting} and \textit{news selecting} rather than in-depth and concentrated reading in current screen-based news consuming processes. 
	In this manner, as shown in Figure \ref{fig.intro}, we scrutinize the news selection behavior and summarize it into three successive steps: (1) \underline{Scan \& Focus}. On spotting entities during \textbf{scanning}, the reader quickly associates them to politics and focuses on \textit{Musk}; 
	(2) \underline{Arouse Interest}. This initial focus stimulates the reader to \textbf{read the second title} intensively, and extends the primitive focus into curiosity and questions (\textit{e.g., Why would Musk leave DOGE?}); 
	(3) \underline{Make Click Decisions}. The reader then decides if the interest is strong enough to be worth a \textbf{click}.
	Adhering to these steps, we find that at the intra-news level, there's a leading entity (\textit{e.g., Musk}) that would be the most attractive one during scanning;	other assistant entities (\textit{e.g., DOGE}) would help to emphasize the interest in this leading entity. 
	Furthermore, at the inter-news level, there's an inter-news entity guided interest stream (\textit{e.g., Trump → Musk}) that will influence title reading. 
	If readers do not have interest in a particular entity, they might not proceed to read the associated title; consequently, clicking becomes more improbable. 
	In other words, the interest of entities at both levels can largely guide the news item click decision. 
	Unfortunately, current methods neither pay attention to the intra-news entity interest in the \textit{scanning} phase nor utilize the guidance from entities at the inter-news level in \textit{title reading} and \textit{clicking}, leading to their failure in achieving optimal performance.
	
	In this work, based on these observations, we devise a novel model called \textbf{IP2} (entity-guided \textbf{I}nterest \textbf{P}robing for \textbf{P}ersonalized news recommendation), to further probe and utilize entity-guided news selection interest.
	At the intra-news level, considering that the news title itself implies the relative importance of each entity, we rely on self-supervised learning to probe informative entities \cite{nishikawa2022ease}. 
	Specifically, a Transformer-based encoder is devised to aggregate entities mentioned in one news story into a single signature entity embedding. 
	Then, a signature entity-title contrastive pre-training is adopted to probe which entity may act as the leading role within one news article. 
	The signature entity provides a unified representation from the entity side for one news item, while the self-supervised learning makes our IP2 feasible to probe intra-news entity interest without requiring interaction logs.
	As for the inter-news level, to highlight the guidance from entities in the news selection chain, a dual-tower user encoder is presented to capture entity-guided and semantic-guided preference streams simultaneously. 
	Furthermore, to probe whether the primitive interest in entities will stimulate a strong enough curiosity upon title reading, and emphasize that title reading could be affected by inter-news entity interest, we further adopt cross attention mechanism between these two towers to ensure an aligned and accurate interest modeling.
	Finally, a learnable aggregation layer is adopted to adjust the importance of entity guidance in news recommendations more personally.
	
	In summary, the contributions of this work are as follows: 
	\begin{itemize}
		\item In this work, we re-summarize news selection behavior into three successive steps: \textit{scanning, title reading}, then \textit{clicking}. 
		We highlight the guidance role of entities on reader's interest modeling among these steps at intra- and inter-news levels.
		
		\item At intra-news level, IP2 utilizes entity-title contrastive pre-training to probe leading entities during \textit{scanning}; at inter-news level, IP2 adopts cross attention to calibrate \textit{title reading} and the final \textit{click decision} via inter-news entity guidance.
		
		\item To the best of our knowledge, we are the first to propose the entity-title contrastive pre-training framework in recommendation. Empirical results on two real-world datasets show that by modeling two levels of entity-guided interest, IP2 can achieve state-of-the-art performance.
	\end{itemize}
	
	\section{Related Work}
	In this section, we briefly discuss two genres of news recommendation methods: neural news recommendation and knowledge-aware neural news recommendation.
	
	\subsection{Neural News Recommendations}
	To mitigate media information overload, news recommender systems are widely studied and adopted in real-world online platforms. 
	Initially, deep factorization machine-based methods \cite{guo2017deepfm, rendle2012factorization} were widely used to generate personalized news feeds based on the user-news consumption matrix.
	With the rapid development of deep learning, recent models have shifted towards the neural collaborative filtering \cite{he2017neural}, which employs deep neural networks to extract news and user features, thereby replacing traditional matrix factorization techniques \cite{guo2017deepfm, rendle2012factorization}. 
	Under this setting, many methods \cite{xiao2023improving} employ static word vectors to encode the news article\footnote{In this work, when referring to a ``news article'', unless explicitly stated otherwise, only the title is considered.}; other models \cite{ma2023punr, kannen2024efficient} find the pretrained language models (PLM) like BERT \cite{kenton2019bert} is more powerful in extracting news features. For user feature extraction, various neural networks are proposed to encode the news reading sequence, such as Recurrent Neural Network (RNN) \cite{an2019LSTUR}, Transformer \cite{zhang2021unbert, ma2023punr} and manually designed attention network \cite{wu2019NRMS, qi2021hierec}. 
	Additionally, various training techniques like Self-Supervised Learning (SSL) \cite{ma2023punr, xiao2023improving} and Information Bottleneck (IB) \cite{xiao2023improving} are adopted to further enhance user feature extraction.
	
	\subsection{Knowledge-aware News Recommendation}
	Normal neural news recommendations are solely based on semantic information.
	Considering that entities mentioned in news articles can link to knowledge graph (KG) nodes, it is natural to utilize entity representations learned from knowledge graphs as external support on top of regular neural news recommenders. To achieve this, there are two lines of research. 
	The first is to enhance the news representations \cite{qi2021hierec, qi2021KIM}. For example, DKN \citep{wang2018dkn} employs a knowledge-aware CNN to generate news representations; KRED \cite{liu2020kred} employs a knowledge graph attention to enhance article representation.
	The second is to enhance the user representations \cite{Chen2024NRMGNR, qiu2022graph}. For instance, PerCoNet \cite{liu2023perconet} adopts an explicit persona analysis based on entities to further understand reading preference; FUM \cite{Qi2022FUMFA} adopts entities accompanied by multi-field attention for fine-grained word-level news interaction modeling.
	There are also attempts using entities as a bridge to jointly enhance news and user representations. For instance, GLORY \cite{yang2023GLORY} incorporates global and local graphs to utilize entities as the proxy for refined user and news story representation learning.
	
	However, the aforementioned two groups of methods tend to prioritize complex models for capturing contextual information to understand a user's needs. 
	In other words, they may not align with real-life news selection behavior and lack a comprehensive understanding of how entity guidance works in news recommendation. Furthermore, KG itself may suffer from entity missing \cite{pujara2017sparsity} and information expiring \cite{leblay2018deriving} issues since KGs can hardly keep up with emerging news events.
	In contrast, our IP2 exploits entity guidance that is in line with real-world behavior and originates entity knowledge from in-context news articles to ensure an accurate and interpretable news recommendation without relying on KG.
	
	\section{Problem Formulation}
	\paragraph{Symbol description.} 
	Let $\mathit{U} = \{ u_1, u_2, \cdots, u_{|\mathit{U}|}\}$ represent the set of all users, and $\mathit{N} = \{ n_1, n_2, \cdots, n_{|\mathit{N}|}\}$ represent the set of all news. In addition, an element in $\mathit{N}$ is a tuple $n_i = \left(t_i, e_i \right)$ where $t$ and $e$ represent for title and entities respectively. Specifically, $e$ is a list of entities $[e_{i1}, e_{i2}, \cdots, e_{ik}]$ that maybe zero length, and $t$ is a list of tokens $[w_{i1}, w_{i2}, \cdots, w_{im}]$. 
	\paragraph{Problem statement.}
	News recommendation is considered as a click prediction problem. 
	Using a list of news to represent users' reading history $hist = [n_1^u, n_2^u, \cdots, n_l^u]$, a news recommender will first encode news $n_i$ into an embedding $n_i'$, then employ a user encoder to aggregate the history sequence into a user embedding $u'$. The preference score of user $u$ about the candidate news $n_c$ will be estimated as the embedding similarity: $\text{sim}(u', n_c')$. Given that the body of news articles can only be viewed after clicking through, it is important to note that in this work, the article bodies will not be used in any of our experiments, unless explicitly stated otherwise.
	
	\section{The Proposed Method}
\begin{figure*}[!t]
		\centering
		\includegraphics[width=0.9\textwidth]{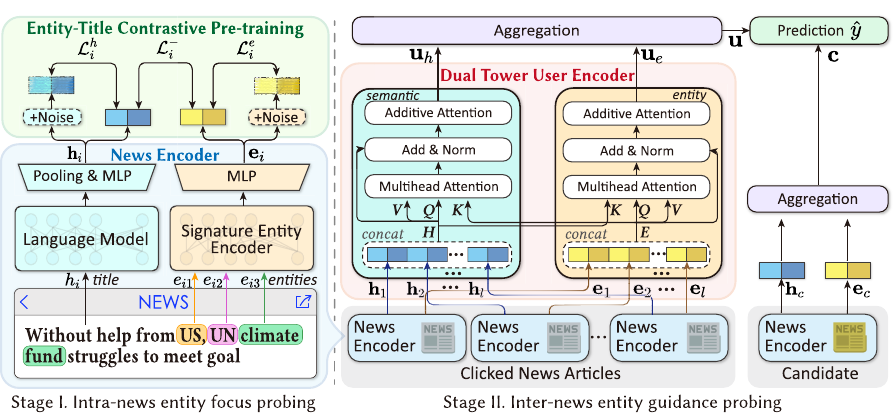}
		\caption{The framework of IP2. IP2 follows a two-stage training paradigm. In the first stage, we conduct signature entity-title contrastive pre-training to probe entity interest at the intra-news level. In the second stage, we employ an entity \& semantic dual tower user encoder to capture entity-guided interest at the inter-news level. These two stages share the same news encoder.} 
		\label{fig.main}
	\end{figure*}

	In this section, we present the proposed IP2 model. The overall model framework is shown in Figure \ref{fig.main}. Different from other methods, IP2 follows a two-stage training paradigm: \textbf{i}. Contrastive pre-training; \textbf{ii}. Downstream news recommendation. 
	\subsection{News Encoder}
	In the context of neural news recommendation, the news encoder aims to extract features from news. In IP2, in order to probe entity-guided personalized news preference, we further incorporate an entity encoder alongside the text-based title encoder.
	
	\subsubsection{Title Encoder}
	The news title encoder aims to obtain basic semantic features from news titles. In this work, to capture in-depth fine-grained semantics information, we adopt the pre-trained BERT as the title encoder. Given title $h$ that contains a list of tokens $[w_1, w_2, \cdots, w_m]$ with maximum length $m$, we feed $h$ into the BERT and then acquire the last layer hidden matrix $\mathbf{H} \in \mathbb{R}^{m \times d}$, where $d$ is the dimension of token embeddings.
	To emphasize the unique contribution of each word, we employ attention pooling to generate the final title embedding. More specifically,
	we first employ a multi-head self-attention layer with $n$ heads to capture intra-news token relations, which is demonstrated as follows:
	\begin{equation}
		\mathsf{MHAttention}(\mathbf{Q}, \mathbf{K}, \mathbf{V}) = \left[ head_1; \dots; head_n \right] \mathbf{W}^O,
	\end{equation}
	\begin{equation}
		head_i = \mathsf{att} ( \mathbf{QW}_i^Q, \mathbf{KW}_i^K, \mathbf{VW}_i^V ), \\
	\end{equation}
	\begin{equation}
		\mathsf{att}(\mathbf{Q}, \mathbf{K}, \mathbf{V}) = \mathsf{softmax}(\mathbf{QK}^\top/\sqrt{d_k})\mathbf{V},
	\end{equation}
	where $\mathbf{W}_i^Q, \mathbf{W}_i^K, \mathbf{W}_i^V $ and $\mathbf{W}^O $ are all learnable parameters.
	To this point, we acquire an intra-news token relationship enhanced embedding matrix $\mathbf{H}' = \mathsf{MHAttention} (\mathbf{H}, \mathbf{H}, \mathbf{H})$. To avoid information loss and noise propagation, we also apply residual link with layer normalization \cite{ba2016layer},
	\begin{equation}
		\widetilde{\mathbf{H}'} = \textsf{layernorm}(\mathbf{H}+\mathbf{H}'),\\
	\end{equation}
	where $\widetilde{\mathbf{H}'}$ can be viewed as a concatenation of multiple token embeddings $[ \widetilde{\mathbf{h}_1'}; \dots ;\widetilde{\mathbf{h}_m'} ]$. To value the importance of different tokens (including entity tokens), we then exploit additive attention to aggregate token embeddings into a title embedding. The attention weight $\alpha_i$ for the $i$-th token embedding is defined as:
	\begin{equation}
		\begin{gathered}
			\alpha_i = \frac{\exp (a_i)}{\sum_{j=1}^{m} \exp (a_j)},\\
			a_i = \mathbf{W}^{(2)}\tanh (\mathbf{W}^{(1)} \widetilde{\mathbf{h}_i'} + \mathbf{b}^{(1)}) + b^{(2)},
		\end{gathered}
	\end{equation}
	where $\mathbf{W}^{(1)}, \mathbf{b}^{(1)}, \mathbf{W}^{(2)} $ and $b^{(2)}$ are all parameters to learn. Through weighted sum, we acquire the final title embedding,
	\begin{equation}
		\mathbf{h} = \mathsf{Add.Attention}(\widetilde{\mathbf{H}'}) =  \sum_{j=1}^{m} \alpha_j \widetilde{\mathbf{h}_j'}.
	\end{equation}
	Notably, unlike the common way to utilize PLM as the sentence encoder, we do not employ the \verb|[CLS]| output to represent a news title. By doing so, we aim to preserve sentence and token-level semantic information simultaneously, which may facilitate intra-news level entity interest probing.
	
	\subsubsection{Entity Encoder}
	The entity encoder aims to extract news features from the entity side. As aforementioned, the leading entity evokes the primary interest during \textit{scanning}, while other entities support and amplify this initial interest. 
	Based on these observations, we introduce the Signature Entity Encoder (SEE), which employs multiple bidirectional Transformer layers to capture each entity's contextual importance by considering all co-occurring entities within one title, allowing SEE to dynamically assign attention weights to pivotal ones.
	
\begin{figure}[!t]
		\centering
		\includegraphics[width=0.9\columnwidth]{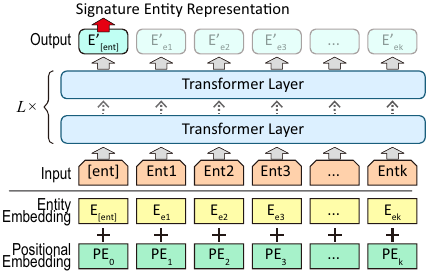}
		\caption{The architecture of Signature Entity Encoder.} 
		\label{fig.entenc}
	\end{figure}

	The architecture of SEE is shown in Figure \ref{fig.entenc}. We first employ a learnable entity memory, denoted as $\mathbf{M} \in \mathbb{R}^{d_c\times d_e}$, to cast entities into the latent space, where $d_c$ is the number of entities in the dataset while $d_e$ is the embedding dimension. 
	For one news article $n = (t, e)$, all the entities mentioned are converted into latent embeddings $\mathbf{E} = \left [E_{[\text{ent}]}, E_{e1}, \dots, E_{ek} \right]$. Inspired by BERT \cite{kenton2019bert}, we also prepend a handle entity \verb|[ent]|.
	We then employ positional embeddings (PE) to preserve the entity presence order, since the position may also affect the entity attention, 
	\begin{equation}
		\label{eq.pe}
		\begin{aligned}    
			\begin{cases}       
				PE_{\mathsf{pos}, 2i} &= \sin(\mathsf{pos}/10000^{2i/d_e})\\
				PE_{\mathsf{pos}, 2i+1}&=\cos(\mathsf{pos}/10000^{2i/d_e})
			\end{cases} 
		\end{aligned},
	\end{equation}
	where $\mathsf{pos}$ is the position, $i$ is the offset inside dimension $d_e$.
	Finally, we stack \textit{L} Transformer layers $\mathsf{Trm}(\cdot)$ on top of each other, to encode the whole input embedding matrix, 
	\begin{equation}
		\mathbf{E}' = \underbrace{\mathsf{Trm}(\dots \mathsf{Trm}}_{L \times}(\mathbf{E} + \mathbf{PE})),
	\end{equation}
	where $\mathbf{E}' \in \mathbb{R}^{k \times d_e}$. We utilize the \verb|[ent]| position output $\mathbf{e} = \mathbf{E}'_{[\mathsf{ent}]}$ as the final signature entity representation for news article $n$. 
	
	Specifically, we would like to point out that the news encoder in IP2 will output a two-element tuple $(\mathbf{h}, \mathbf{e})$ for a given input news article $n$. Both title and entity encoding procedures remain the same in contrastive pre-training and the downstream recommendation. 
	
	\subsection{Entity-Title Contrastive Pre-training}
	To this point, IP2 is ready to extract features from the semantic and entity sides. 
	However, we build the SEE from scratch, which means knowledge is not included at this moment.
	Unlike other knowledge-aware methods, we do not incorporate external knowledge graphs as the knowledge source; the considerations are twofold: (1) The knowledge graph itself has the sparsity issue \cite{pujara2017sparsity}, which means there are always newly emerged entities that can not be linked to the knowledge graph. (2) Learning knowledge graph embedding is time-consuming, and may easily suffer from information expiring issue since KGs can hardly keep up with rapid changes in news stories \cite{leblay2018deriving}.
	Considering that at the word level, PLM has an inherent attention on different tokens (including entities), at the news article level, PLM can describe an entity from multiple angles. Inspired by \citet{nishikawa2022ease}, we employ a signature entity-title contrastive pre-training, which can quickly initialize the entity memory $\mathbf{M}$ with the proper meaning, and adapt token attention into intra-news entity interest.
	
	Taking a mini-batch of $v$ news articles $[n_1, n_2, \dots, n_v]$ as an example. For article $n_i$, we employ the aforementioned news encoder to derive title embedding $\mathbf{h}_i$ and signature entity embedding $\mathbf{e}_i$. We then employ MLP layers to cast them into an identical dimension size $\mathbb{R}^d$. With acquired news title embeddings $[\mathbf{h}_1, \mathbf{h}_2, \dots, \mathbf{h}_v]$ and signature entity embeddings $[\mathbf{e}_1, \mathbf{e}_2, \dots, \mathbf{e}_v]$, we choose a group of embedding pairs $\mathcal{P}=\{ (\mathbf{e}_i, \mathbf{h}_i)\}_{i=1}^v$ as the positive pair set, and the opposite scenario $\mathcal{N} =\{ (\mathbf{e}_i, \mathbf{h}_j)\}, i\ne j$ as the negative pair set. Following \citet{pmlr-v119-chen20j}, the training loss for positive pairs is defined as follows:
	\begin{equation}
		\mathcal{L}_i^- = - \log \frac{\exp \left(\mathsf{sim} (\mathbf{e}_i,\mathbf{h}_i)  / \tau\right) }{\sum_{j=1}^{v} \exp(\mathsf{sim} (\mathbf{e}_i,\mathbf{h}_j)/ \tau) },
		\label{eq.cl_raw}
	\end{equation}
	where $\tau$ is the temperature, \textsf{sim} is the cosine similarity.
	Contrastive learning aims to pull positive samples together and push negative samples away \cite{wang2021understanding, zhang2025rethinking}. To avoid these two types of embeddings becoming homogenized, we follow \citet{inoue2019multi} and apply dropout layers as noisy gates to slightly modify an embedding into a ``mirror embedding'', then make every two of them the positive pair and modify (\ref{eq.cl_raw}) into:
	\begin{equation}
		\mathcal{L}_i^h = - \log \frac{\exp \left(\mathsf{sim} (\mathbf{h}_i,\mathbf{h}_i^+)  / \tau\right) }{\sum_{j=1}^{v} \exp(\mathsf{sim} (\mathbf{h}_i,\mathbf{h}_j^+)/ \tau) },
	\end{equation}
	in which $\left(\mathbf{h}_i,\mathbf{h}_i^+ \right)$ stands for the original title embedding and its perturbed mirror. The same technique also applies to entities and yields another loss function,
	\begin{equation}
		\mathcal{L}_i^e = - \log \frac{\exp \left(\mathsf{sim} (\mathbf{e}_i,\mathbf{e}_i^+)  / \tau\right) }{\sum_{j=1}^{v} \exp(\mathsf{sim} (\mathbf{e}_i,\mathbf{e}_j^+)/ \tau) }.
	\end{equation}
	
	The overall contrastive learning loss is defined as follows, in which $\alpha$, $\beta$, and $\delta$ are all hyperparameters, sum up to 1.
	Specifically, these three parts can have their own temperature $\tau$.
	\begin{equation}
		\mathcal{L}_{i} = \alpha \mathcal{L}_{i}^- + \beta\mathcal{L}_{i}^h + \delta\mathcal{L}_{i}^e.
		\label{eq.clloss}
	\end{equation}
	
	The benefits of utilizing contrastive pre-training to probe intra-news entity interest are twofold: First, through contrastive pre-training, we can learn entity embeddings (stored in entity memory $\mathbf{M}$) without relying on knowledge graphs, which makes our IP2 still functional when KG is not available. 
	Second, the self-supervised contrastive pre-training enables IP2 to harness numerous news articles for intra-news entity interest probing without requiring interaction logs.
	
	\subsection{Dual Tower User Encoder}
	The user encoder plays a crucial role in extracting features from a user's news reading history. In this section, we shed light on the use of entity guidance in addition to semantic information to accurately capture and model the user's reading preferences.
	
	\subsubsection{Dual Tower User Encoder with Cross Attention}
	As mentioned earlier, the interest between entities at the inter-news level can play a crucial role in guiding the entire news selection process. In our IP2, we aim to utilize this guidance signal along with the semantic information as two interest streams.
	To achieve this, we base our model on an attention-based user encoder \cite{wu2019NRMS, wu2021PLMNR} with two identical attention towers in parallel. The first tower focuses on probing reading interest based on semantics, while the second tower focuses on entities.
	In addition, considering that the initial entity interest will stimulate the user to read the title intensively, which may, in turn, spark curiosity about other entities. In other words, there is an interaction between these two streams. Thus, we take inspiration from the modal-wise interaction in vision-language models \cite{lu2019vilbert} and incorporate a cross attention mechanism to model this stream-wise interest fluctuation.
	
	Given a sequence of title embeddings $\mathbf{H} = [ \mathbf{h}_1, \mathbf{h}_2, \dots, \mathbf{h}_l]$ and entity embeddings $\mathbf{E} = [ \mathbf{e}_1, \mathbf{e}_2, \dots, \mathbf{e}_l]$ from a click history with length $l$, for the semantics-based interest stream, we employ aggregated multi-head cross attention with residual link to make the encoding procedure also aware of entities and reduce information loss. Detailed procedures are described as follows:
	\begin{equation}
		\begin{split}
			\mathbf{U}_h = \mathsf{MH Attention}(\mathbf{H}, \mathbf{E}, \mathbf{H}),\\
			\widetilde{\mathbf{U}_h} = \mathsf{layernorm} (\mathbf{H} + \mathbf{U}_h),\\
			\mathbf{u}_h = \mathsf{Add. Attention}(\widetilde{\mathbf{U}_h}).
		\end{split}
	\end{equation}
	
	Notably, we incorporate the title and entity embedding matrix as the \textit{query} and \textit{key}, respectively. 
	For the entity-based preference stream, we also adopt the same architecture,
	\begin{equation}
		\begin{split}
			\mathbf{U}_e = \mathsf{MH Attention}(\mathbf{E}, \mathbf{H}, \mathbf{E}),\\
			\widetilde{\mathbf{U}_e} = \mathsf{layernorm} (\mathbf{E} + \mathbf{U}_e), \\
			\mathbf{u}_e = \mathsf{Add. Attention}(\widetilde{\mathbf{U}_e}),
		\end{split}
	\end{equation}
	where we utilize entity and title embedding matrix as the \textit{query} and \textit{key}, respectively.  
	
	The merits of using a dual tower user encoder with cross attention are threefold: 
	First, using two towers prevents the over-emphasizing of news articles with specific semantics or entities, thereby maintaining diversity in the recommendation results.
	Second, the attention mechanism is well-suited for capturing both long-term and short-term interest within a user's click sequence. 
	Third, cross-tower attention enables modeling stream-wise interactions, allowing for a more dynamic understanding of users' interest.
	
	\subsubsection{Aggregation}
	Based on the aforementioned title-based and entity-based encoder towers, we obtain two user preference embeddings $\mathbf{u}_h$ and $\mathbf{u}_e$. Considering that different users may balance these two preference streams differently, in other words, while most users may directly skip title reading if they are not interested in the associated entities, there are users who may still take a glance. In light of this phenomenon, we employ a weighted sum of $\mathbf{u}_h$ and $\mathbf{u}_e$ as the final user preference embedding $\mathbf{u}$,
	\begin{equation}
		\begin{split}
			\mathbf{u} = \eta_h\mathbf{u}_h + (1-\eta_h)\mathbf{u}_e, \\
			\eta_h = \sigma([\mathbf{u}_h;\mathbf{u}_e]\mathbf{W_a} + b_a),
		\end{split}
		\label{eq.agg1}
	\end{equation}
	where $\mathbf{W_a}$ and $b_a$ are parameters to learn. We can also acquire two embeddings for one candidate news article: title embedding $\mathbf{h}_c$ and entity embedding $\mathbf{e}_c$. We take the same weighted sum method to aggregate these two embeddings and get the final candidate news embedding $\mathbf{c}$,
	\begin{equation}
		\begin{split}
			\mathbf{c} = \eta_c\mathbf{h}_c + (1-\eta_c)\mathbf{e}_c, \\
			\eta_c = \sigma([\mathbf{h}_c;\mathbf{e}_c]\mathbf{W_a} + b_a).
		\end{split}
		\label{eq.agg2}
	\end{equation}
	
	\subsection{Model Training}
	The proposed IP2 model follows a two-stage training strategy. The first stage, which we refer to as ``pre-training'', focuses on entity-title contrastive learning. During this stage, the optimizing target is given by equation (\ref{eq.clloss}), and we will export the news encoder into a checkpoint at the end of the last epoch.
	
	The second stage involves training the model for the regular recommendation task. At this stage, we use negative sampling to choose one positive news (clicked) $n_i^+$ and $r$ negative news (not clicked) $[n_i^{1-}, n_i^{2-},\cdots, n_i^{r-}]$ within the same \textit{i}-th session. We first initialize the model using the aforementioned checkpoint, then utilize the widely adopted \cite{yang2023GLORY, rendle2020neural} dot product to calculate the click probability score $\hat{\mathbf{y}}_i = [\hat{y}_i^{+}, \hat{y}_i^{1-}, \hat{y}_i^{2-},\cdots, \hat{y}_i^{r-}]$ for each news article,
	\begin{equation}
		\hat{\mathbf{y}}_i = \mathsf{softmax}(\mathbf{u} \cdot \mathbf{c_i}),
	\end{equation}
	where $\mathbf{c_i}$ contains one positive and $r$ negatively sampled news embeddings in \textit{i}-th session. Finally, we optimize the log-likelihood loss $\mathcal{L}_{NCE}$ for all positive samples over the entire training set $\mathcal{S}$.
	\begin{equation}
		\mathcal{L}_{NCE} = -\sum_{i=1}^{|\mathcal{S}|}\log \frac{\exp (\hat{y}_i^+)}{\exp (\hat{y}_i^+) + \sum_{1}^{r}\exp (\hat{y}_i^{j-})}.
		\label{eq.celoss}
	\end{equation}
	Notably, loss functions in these two stages are independent. Compared to the commonly employed cross-entropy loss, utilizing NCE loss enables IP2 to effectively leverage additional information derived from negative feedback. 
	
	\section{Experiments} 
	In this section, we conduct experiments to evaluate the performance of our IP2, and shed light on these key Research Questions:
	\begin{itemize}
		\item\textbf{RQ1:} Does the proposed IP2 surpass state-of-the-art news recommendation baseline methods?
		\item	\textbf{RQ2:} How does the signature entity-title contrastive pre-training help to probe intra-news level entity interest?
		\item	\textbf{RQ3:} Does the specially designed user encoder utilize inter-news entity guidance better in interest modeling?
		\item \textbf{RQ4:} What is the influence of the PLM size and the sizes of other components in IP2?
		\item \textbf{RQ5:} How does IP2 perform in real-world instances?
	\end{itemize}

	\subsection{Experimental Setup}
	\subsubsection{Datasets and Preprocessing}
	We conduct extensive experiments on two real-world datasets. One is MIND \cite{wu-etal-2020-mind}, which was collected from 6 weeks of anonymized behavior logs of the MSN News website. 
	We utilize both the full \textit{large} version and the sampled \textit{small} version of MIND.   
	The other is Adressa-\textit{1week} \cite{gulla2017adressa} released by the Norwegian newspaper company Adresseavisen. For Adressa, following previous works \cite{yang2023GLORY, liu2023perconet}, we build the training and testing sets using logs from the first 6 days and the last day, respectively.
	Since Adressa does not provide impression lists that contain negative samples, for each click, we randomly sample 20 news articles for testing.
	Detailed statistics can be found in Table \ref{table.sta}.
	For data preprocessing, we drop all reading logs that are shorter than 5, truncate the logs that are longer than 50, and limit the title length to 20 words.
	
	\subsubsection{Implementation Details.}
	We build the title encoder based on BERT-Base\footnote{We use the Norwegian BERT \textsf{NbAiLab/nb-bert-base} on Adressa.}, while the signature entity encoder contains \textit{L}=2 Transformer layers is built from scratch. For model training, we utilize the AdamW optimizer with initial learning rates $\gamma = 1e^{-5}$ for BERT and $\gamma = 1e^{-4}$ for non-BERT parts that are linearly decayed with 10\% warm-up steps. Our IP2 follows a two-stage training strategy. We employ the same $\tau=0.1$ in the first two-epoch  \textit{contrastive pre-training stage} with $\alpha$, $\beta$, and $\delta$ set to 0.3, 0.2, and 0.2, respectively, while the batch size is set to 128. 
	In the second \textit{downstream recommendation stage} we use (\ref{eq.celoss}) as the optimization target with $r=4$ negative samples, and the batch size is set to 64.
	We utilize the official Microsoft recommenders library\footnote{\url{https://github.com/microsoft/recommenders}} or the source code provided by the original authors to build all baseline models. 
	Following the original settings used in MIND \cite{wu-etal-2020-mind}, we use AUC, MRR, nDCG@5, and nDCG@10 as evaluation metrics. 
	All models are trained on one NVIDIA A100-SXM4-80GB GPU. The source code repository is available at \url{https://doi.org/10.5281/zenodo.15861071}.
	\begin{table}[!t]
		\caption{Dataset Statistics.}
		\begin{tabular}{lrrr}
			\toprule
			&MIND-small& MIND-large&Adressa\\
			\midrule
			\#users&  94,057 & 1,000,000 & 640,503\\
			\#news&   65,238 & 161,013 & 20,428\\
			\#entities &31,451 & 42,628 & 98,596\\
			\#clicks & 347,727 & 24,155,470 & 3,101,991\\
			\#impressions & 230,117 & 15,777,377 & -\\
			\bottomrule
		\end{tabular}
		\label{table.sta}
	\end{table}	
	\subsubsection{Compared Methods.}
	We take the following two groups of state-of-the-art methods as the baselines.
	
	\textbf{Neural News Recommendation Methods}: 
	(1) \textbf{NRMS} \cite{wu2019NRMS} applies multi-head self-attention to learn news and user representations; 
	(2) \textbf{GERL} \cite{ge2020graph} utilizes a news-user bipartite graph to better capture high-order relatedness between users and news; 
	(3) \textbf{HieRec} \cite{qi2021hierec} adopts hierarchical structure to capture users' diverse and multi-grained
	interest in multiple levels;
	(4) \textbf{PLM-NR} (also known as NRMS-BERT) enhances NRMS by using off-the-shelf PLM for news feature extraction; 
	(5) \textbf{UNBERT} \cite{zhang2021unbert} leverages PLM to capture multi-grained user-news matching signals at both word-level and news-level; 
	(6) \textbf{DIGAT} \cite{mao2022digat} utilizes dual-graph interaction between user-user and news-news graphs for accurate news-user matching; 
	(7) \textbf{PUNR} \cite{ma2023punr} incorporates PLM-inspired pre-training tasks for enhanced user interest modeling;
	(8) \textbf{TDNR-C\textsuperscript{2}} \cite{shu2024don} uses contrastive learning to mitigate content authenticity bias.
	
	\begin{table*}[t]
		\caption{Results on MIND-small and MIND-large. The best results are marked in boldface, while the second-best results are highlighted using underlines. ``\dag'': results taken from \cite{liu2023perconet}, dash (-) means the result cannot be obtained due to source code unavailability; ``*'': improvements are significant at the level of 0.05 with paired \textit{t}-test.}
		\label{table.result}
		\centering
		\begin{center}
			\begin{tabular}{rcccccccc}
				\toprule
				\multirow{2}{*}{\textbf{Method}} & \multicolumn{4}{c}{\textbf{MIND-small}} & \multicolumn{4}{c}{\textbf{MIND-large}} \\ 
				\ifpdf
				\cmidrule(lr){2-5}
				\cmidrule(lr){6-9}
				\else
				\cline{2-9}
				\fi
				& AUC & MRR & nDCG@5 & nDCG@10 & AUC & MRR & nDCG@5 & nDCG@10  \\
				\midrule
				NRMS & 65.63 & 30.96 & 34.13 & 40.52 & 68.24 &  33.49 & 36.56 & 42.24 \\
				GERL & 65.27 & 30.10 & 32.93 & 39.48 & 68.10 & 33.41 & 36.34 & 42.03 \\
				HieRec & 67.95 & 32.87 & 36.36 & 42.53 & 69.03 & 33.89 & 37.08 & 43.01 \\
				PLM-NR & 68.60 & 32.97 & 36.55 & 42.78 & 69.50 & 34.75 & 37.99 & 43.72 \\
				UNBERT & 67.92 & 31.72 & 34.75 & 41.02 & 70.68 & \underline{35.68} & \underline{39.13} & 44.78 \\
				DIGAT & 68.77 & 33.46 & 37.14 & \underline{43.39} & 70.08 & 35.20 & 38.46 & 44.15 \\
				PUNR & 68.89 & 33.33 & 36.94 & 43.10 & \underline{71.03} & 35.17 & 39.04 & \underline{45.41} \\
				TDNR-C\textsuperscript{2} & 68.89 & 33.57 & 37.23 & \underline{43.39} & 70.38 & 34.62 & 38.12 & 44.30 \\
				\midrule
				DKN & 62.90 & 28.37 & 30.99 & 37.41 & 64.07 & 30.42 & 32.92 & 38.66 \\
				KRED & 65.33 & 30.60 & 33.42 & 39.98 & 68.52 & 33.78 & 36.76 & 42.45 \\
				UaG & 65.10 & 29.89 & 33.31 & 39.46 & 69.23 & 34.14 & 37.21 & 43.04 \\
				GREP & 68.12 & \underline{33.75} & \underline{37.25} & 43.37 & 69.44 & 34.40 & 37.54 & 43.22 \\
				FUM & 67.11 & 31.31 & 35.08 & 41.42 & 70.01 & 34.51 & 37.68 & 43.38 \\
				PerCoNet\textsuperscript{\dag} & \underline{68.93} & 33.40 & 36.93 &43.28 & - & - & - & - \\
				GLORY & 68.15 & 32.97 & 36.47 & 42.78 & 69.45 & 34.03 & 37.92 & 44.19 \\
				\midrule
				\textbf{IP2}& \textbf{69.69}* & \textbf{34.51}* & \textbf{38.30}* & \textbf{44.42}* & \textbf{72.06}*&\textbf{35.96}*&\textbf{40.09}*&\textbf{46.35}* \\
				\bottomrule
			\end{tabular}
		\end{center}
		
	\end{table*}
	\textbf{Knowledge-aware Neural News Recommendation Methods}: 
	(1) \textbf{DKN} \cite{wang2018dkn} uses a knowledge-aware CNN to fuse semantic-level and knowledge-level representations of the news; 
	(2) \textbf{KRED} \cite{liu2020kred} devises a knowledge-aware GNN to learn news representations from news titles and entities; 
	(3) \textbf{User-as-Graph} (UaG) \cite{wu2021UAG} learns user interests via heterogeneous graph pooling on personalized graphs;
	(4) \textbf{GREP} \cite{qiu2022graph} incorporates knowledge graph convolution and news-entity bipartite graph to capture existing and potential reading interest; 
	(5) \textbf{FUM} \cite{Qi2022FUMFA} leverages entities as interest clues in news selection by incorporating a multi-document Fastformer architecture; 
	(6) \textbf{PerCoNet} \cite{liu2023perconet} adopts prominent entity-based explicit persona analysis for explainable user representation learning; 
	(7) \textbf{GLORY} \cite{yang2023GLORY} combines global and local news and entity graphs to enhance news reading behavior modeling in different contexts.
	
	\subsection{Overall Performance}
	The overall performance is shown in Table \ref{table.result} and \ref{table.adressa}. We run each experiment 5 times with different random seeds and report averaged results to ensure robustness. Notably, all the numbers listed here are percentage numbers with ``\%'' omitted. Through these results, we have the following observations: 
	
	First, methods that consider fine-grained interest (\textit{e.g.,} HieRec considers topics) perform better than pure text-based methods (\textit{e.g.,} NRMS) since solely relying on semantics is relatively coarse-grained for user modeling. 
	By providing in-depth semantic information, PLMs can boost the performance (\textit{e.g.,} PLM-NR). 
	Incorporating entities provides a different view of user behavior that can yield better results. For instance, PerCoNet utilizes entity-based personality analysis to enhance the user encoder in PLM-NR; GLORY further adopts entity graphs with different contexts to enhance GERL. 
	
	Additionally, we find that utilizing entities does not always perform well. DKN encodes news articles solely based on entities and performs the worst.
	Similar to IP2, UNBERT tries to model intra- and inter-news word-level interest. However, unnecessary words may bring noise that can contaminate the actual reading preference. 
	Moreover, its single-encoder design suffers from the \textit{seesaw issue} that can hardly balance two levels of interest.
	Similarly, FUM tries to model entity-guided interest, but it overlooks the intra-news level.
	While GREP utilizes an entity-dedicated user encoder, its GNN backbone is prone to encountering the cold-start problem on the test set. PUNR achieves remarkable results on MIND-large; however, its BERT-like pre-training task demands a significant amount of interaction logs, making it suboptimal on MIND-small.
	
	\begin{table}[!t]
		\caption{Results on Adressa-1week. ``*'': improvements are significant at the level of 0.05 with paired \textit{t}-test.}
		\label{table.adressa}
		\centering
		\begin{tabular}{rcccc}
			\toprule
			\multirow{2}{*}{\textbf{Method}} & \multicolumn{4}{c}{\textbf{Adressa-1week}} \\ 
			\ifpdf
			\cmidrule(lr){2-5}
			\else
			\cline{2-5}
			\fi
			& AUC & MRR & nDCG@5 & nDCG@10  \\
			\midrule
			NRMS & 75.31 & 42.24 & 44.66 & 48.46\\
			HieRec & 78.67 & 49.22 & 48.72 & 56.67 \\
			PLM-NR & 78.20 & 47.26 & 48.41 &54.60\\
			PUNR & 78.32 & 47.71 & 49.32 & 54.80 \\
			\midrule
			\textbf{IP2}& \textbf{83.16}*& \textbf{50.83}*& \textbf{54.05}* & \textbf{59.32}*\\
			\bottomrule
		\end{tabular}
	\end{table}
	Furthermore, due to the lack of a comprehensive knowledge graph in Norwegian, all knowledge-aware methods that explicitly utilize entity embeddings learned from KG do not work on Adressa-1week\footnote{Some neural NR models may also be inapplicable due to missing metadata. For instance, TDNR-C\textsuperscript{2} requires article abstracts, which are not available in Adressa.}. 
	In contrast, by acquiring entity representations through contrastive pre-training rather than relying on KG, our IP2 is still functional. 
	Unlike experiments on MIND, both PUNR and PLM-NR are inferior to HieRec. 
	This is because limited semantics (average title length, MIND: 10.79 \textit{vs} Adressa: 6.57) restricts interest modeling capability of text-only methods. 
	It further reveals that fine-grained guidance is essential in news recommendations. 
	
	Finally, it is evident that IP2 outperforms all compared methods in all cases (\textbf{RQ1}). 
	IP2 takes the merit of fine-grained entity guidance signal at both intra-news and inter-news levels to overcome shortcomings encountered by other methods. 
	Moreover, benefiting from self-supervised learning, IP2 is less susceptible to the limitations of available interaction data and demonstrates improvements in datasets of varying sizes.
	
	\subsection{Ablation Study}\label{ablation}
	We conduct experiments on MIND-small with the following IP2 variations to evaluate each component's contribution:
	(\textbf{i}) \underline{\textit{w/o} Intra} without intra-news entity interest removes entity-title contrastive pre-training. We also try to partially remove the entity-title part (\underline{\textit{w/o} CL\textsubscript{e-t}}) or both title-title and entity-entity together (\underline{\textit{w/o} CL\textsubscript{t-t\&e-e}}) to evaluate their contributions separately.
	(\textbf{ii}) \underline{\textit{w/o} Inter} without inter-news entity guidance removes cross attention and uses self-attention inside each user encoder tower.
	(\textbf{iii}) \underline{\textit{w/o} Agg} without aggregation layer concatenates $\mathbf{u}_h$ and $\mathbf{u}_e$ in  (\ref{eq.agg1}), $\mathbf{h}_c$ and $\mathbf{e}_c$ in (\ref{eq.agg2}) without using learnable weights.  For each variant, we only make a single modification to the model while keeping other parts intact. All experiments are conducted on MIND-small.
	
	Through results presented in Table \ref{table.ablation}, we find that all proposed components are necessary to improve the performance.
	The model collapses on \underline{\textit{w/o} Intra} and \underline{\textit{w/o} Inter}, confirming the critical role of entity-guided interest probing at both levels (\textbf{RQ2\&3}).
	It is worth noting that \underline{\textit{w/o} CL\textsubscript{e-t}} has severer impacts comparing to \underline{\textit{w/o} CL\textsubscript{t-t\&e-e}}. 
	This reveals that CL\textsubscript{e-t} directly affects whether entity memory $\mathbf{M}$ could be initialized with proper meanings, without which, the SEE may struggle to probe the intra-news level entity focus.
	Besides, \underline{\textit{w/o} Inter} performs the worst, which shows that inter-news stream-wise interest interaction contributes most to the users' reading decision. 
	Finally, the \underline{\textit{w/o} Agg} results suggest that entity- and semantic-guided reading preferences may hold varying importance for different users.
	
	\begin{table}[!t]
		\caption{The ablation results on various IP2 variants. ``\textit{w/o}'' stands for ``without'', ``N@k'' represents ``nDCG@k''.}
		\label{table.ablation}
		\centering
		\begin{tabular}{lcccc}
			\toprule
			\textbf{Variant}&   \textbf{AUC} & \textbf{MRR} & \textbf{N@5}& \textbf{N@10} \\
			\midrule
			IP2 \textit{(original)} & \textbf{69.69}* & \textbf{34.51}* & \textbf{38.30}* & \textbf{44.42}*\\
			\textit{w/o} Intra &  68.69 & 33.52 & 36.98 & 43.29\\
			\quad\textit{w/o} CL\textsubscript{e-t} &  68.63 & 33.70 & 37.16 & 43.37\\
			\quad\textit{w/o} CL\textsubscript{t-t\&e-e} &  69.03 & 34.01 & 37.58 & 43.80\\
			\textit{w/o} Inter &  68.66 & 33.28 & 36.65 & 43.04\\
			\textit{w/o} Agg & 68.44 & 33.38 & 36.87 & 43.12\\
			\bottomrule
		\end{tabular}
		
	\end{table}
	\subsection{Analysis on Contrastive Pre-training (RQ2)}
	
	IP2 is novel in acquiring entity representations through contrastive pre-training, which also makes intra-news level entity focus probing feasible. 
	In this part, we provide a direct comparison between initializing the SEE entity memory $\mathbf{M}$ with \underline{Random} values (IP2's default setup) and \underline{TransE-Wikidata} embeddings, which are utilized by other knowledge-aware methods, to shed light on how signature entity-title contrastive pre-training works in our IP2. 
	
\begin{figure}[!t]
		\centering
		\includegraphics[width=1\columnwidth]{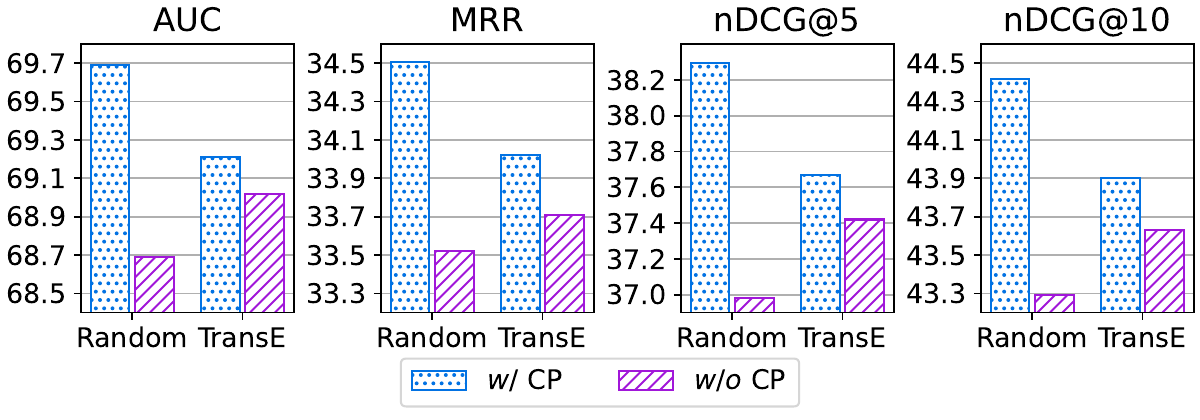}
		\caption{Results on different SEE entity memory setups. ``CP'': signature entity-title Contrastive Pre-training;
			``\textit{w/}'' stands for ``with'' while ``\textit{w/o}'' stands for ``without''. ``Random \textit{w/} CP'' is IP2's default setting.} 
		\label{table.embedding}
	\end{figure}

	As shown in Figure \ref{table.embedding}, \underline{Random \textit{with} CP} performs the best. 
	Additionally, our IP2 still achieves acceptable outcomes with off-the-shelf TransE embeddings even without CP. 
	This can be attributed to the knowledge carried by TransE embeddings continues to be effective in our dual tower user encoder. 
	With contrastive pre-training, the SEE can further probe intra-news level entity interest and achieve improved performance in both settings.
	However, due to the inherent issues with KG, using TranE embeddings may introduce unexpected noise that hinders performance, ultimately leading to suboptimal recommendation results.
	\subsection{Analysis on Model Size (RQ4)}
	\subsubsection{The Impact of PLM Size}
\begin{figure}[!t]
		\centering
		\includegraphics[width=1\columnwidth]{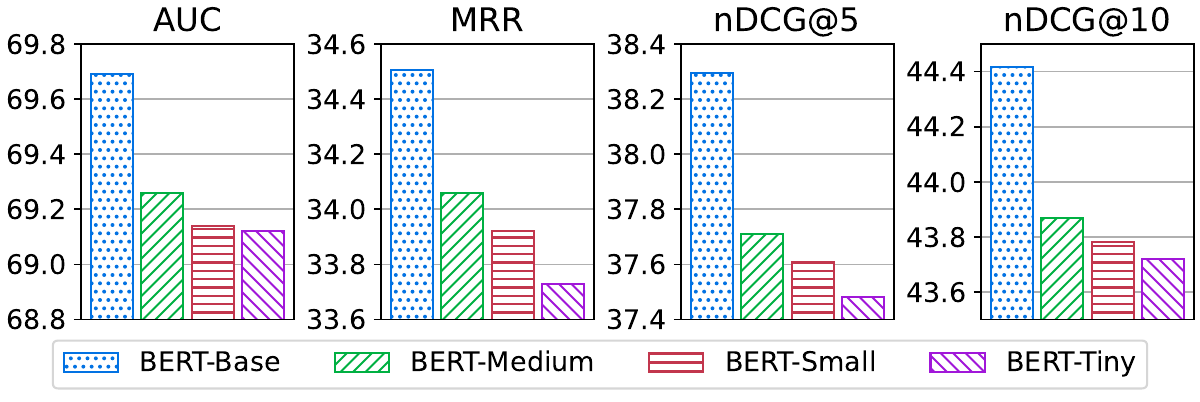}
		\caption{Impact of the BERT size.} 
		\label{fig.bertver}
	\end{figure}

	Since we utilize PLM as the title encoder, we first examine the influence of utilizing different sizes of PLMs, such as BERT-Base, BERT-Medium, BERT-Small, and BERT-Tiny, which consist of 12, 8, 4, and 2 Transformer layers, respectively.
	From the results shown in Figure \ref{fig.bertver}, we find that using larger PLMs usually leads to better performance. 
	This is expected because a larger PLM possesses the capability to capture more detailed semantic information and contains more prior knowledge. These factors can be beneficial for title encoding and entity-title contrastive pre-training, ultimately leading to a better outcome.
	We believe that using 24-layer BERT-Large can further improve the performance. However, this may disrupt the performance-efficiency trade-off, making it less suitable for online applications.
	In addition, it is worth highlighting that even using a moderate BERT-Medium, our IP2 still outperforms all baseline methods.
	
	\subsubsection{The Impact of SEE Size}
\begin{figure}[!t]
		\centering
		\includegraphics[width=1\columnwidth]{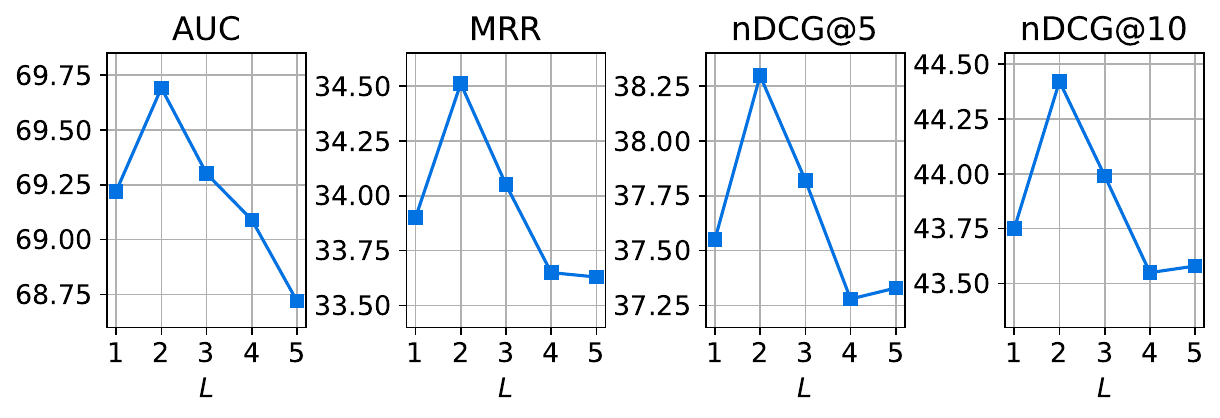}
		\caption{Impact of the SEE size.} 
		\label{fig.seeL}
	\end{figure}

	We then investigate the impact of the hyperparameter $L$ in IP2.
	In particular, we vary $L$ in $\{1,2,3,4,5\}$ and conduct experiments on MIND-small, while keeping other model components unchanged. 
	As shown in Figure \ref{fig.seeL}, the performance initially improves with an increase in \textit{L}, reaching the optimal results at \textit{L}=2, but then declines.
	This observation suggests that with fewer Transformer layers, the SEE may struggle to capture sufficient entity attention information during contrastive pre-training. 
	On the other hand, unlike the PLM utilized in the title encoder, a larger SEE does not necessarily guarantee a better outcome. This could be because a deeper SEE may be prone to overfitting.
	
	
	\subsection{Case Study (RQ5)}
\begin{figure}[!t]
		\centering
		\includegraphics[width=1\columnwidth]{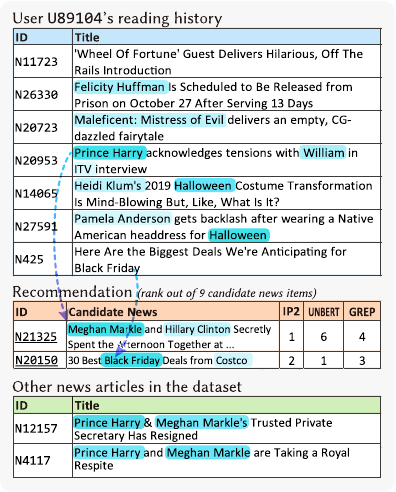}
		\caption{Case study based on a sampled impression log. Entities are highlighted based on attention weights in SEE; darker colors indicate relatively more important. Candidate news items that are clicked by the user are highlighted using underlines.} 
		\label{fig.case}
	\end{figure}

	We further conduct a case study to illustrate IP2's effectiveness in real-world cases.
	As shown in Figure \ref{fig.case}, we take a sampled impression from the interaction log of user \texttt{U89104}, in which the user clicked 2 of the 9 candidate news items after reading 7 articles. Compared to UNBERT and GREP, our IP2 performs the best.
	During contrastive pre-training, since ``Prince Harry'' and ``Meghan Markle'' appear together multiple times (\textit{e.g.,} \texttt{N12157}, \texttt{N4117}) in the dataset, our IP2 recognizes them as a couple within the royal family from the context and memorizes them in the entity memory $\mathbf{M}$. 
	Based on semantics, our IP2 further probes ``Prince Harry'' in \texttt{N20953} as the intra-news entity focus. 
	Equipped with entity-dedicated user encoders, both IP2 and GREP are capable of capturing inter-news entity interest between \texttt{N20953} and \texttt{N21325}.
	However, the literal meaning between \texttt{N20953} and \texttt{N21325} differs so much, which makes GREP struggle between balancing entity and semantics that finally ranks \texttt{N21325} to \#4. In contrast, with cross attention link and aggregation layer, our IP2 is feasible to balance these two aspects, finally ranking \texttt{N21325} to \#1.
	We also find that user \texttt{U89104} is interested in holidays based on reading history. However, ``Black Friday'' is not labeled as an entity\footnote{In this work, we do not perform the named entity recognition (NER) process. We use the entity annotations provided by MIND and Adressa directly. } in \texttt{N425}. 
	For UNBERT, since there is a direct word match, \texttt{N20150} is ranked as \#1. However, this entity missing affects the user encoder in GREP, making \texttt{N20150} being ranked at position \#3. While in our IP2, the cross attention link bridges these two ``black friday''s together, finally ranks \texttt{N20150} to position \#2. 
	\section{Conclusion and Future Work}
	In this work, we scrutinize and summarize the news selection process into three key stages: \textit{scanning}, \textit{title reading}, and \textit{clicking}. We find that intra-news entity interest predominantly influences scanning, whereas inter-news entity-guided interest impacts title reading and the subsequent click decisions.
	Motivated by this observation, we propose IP2 to utilize entities at both levels for a more accurate news recommendation.
	More specifically, IP2 utilizes entity-title contrastive pre-training for intra-news entity interest probing, then employs a cross attention enhanced dual tower user encoder to probe inter-news reading interest.
	Extensive experiments demonstrate that explicitly modeling these two levels of entity-guided interest enables IP2 to achieve state-of-the-art performance. Furthermore, our results highlight the strong capability of language models to understand entities, which can be effectively harnessed in recommendations through a simple contrastive pre-training task. 
	As for future work, we plan to conduct user studies to further reveal the entity-related cognitive steps in online recommendations.
	We also plan to explore the application of this multi-level entity guidance framework in other domains, such as biomedical recommendation.
	\begin{acks}
		We would like to thank our anonymous reviewers for their 
		insightful comments. This work was supported by
		the National Natural Science Foundation of China (No. 62376051, 62076046).
	\end{acks}
	\bibliographystyle{ACM-Reference-Format}
	\bibliography{sample-base}
	
\end{document}